\documentclass[twocolumn,showpacs,amsmath,amssymb]{revtex4}

\usepackage{graphics}
\usepackage[dvips]{graphicx}

\def\sqr#1#2{{\vcenter{\hrule height.#2pt\hbox{\vrule width.#2pt height#1pt
\kern#1pt \vrule width.#2pt}\hrule height.#2pt}}}

\begin{document}

\title{Flat rotation curves using scalar-tensor theories}

\author{Jorge L Cervantes-Cota$^1$, M A  Rodr\'iguez-Meza$^1$ and Dario Nu\~nez$^2$}

\address{$^1$Depto.  de F\'{\i}sica, Instituto Nacional de Investigaciones Nucleares,
A.P. 18-1027, 11801 D.F., M\'exico.  \\
$^2$Instituto de Ciencias Nucleares,  Universidad Nacional
Aut\'onoma de M\'exico, A.P. 70-543,  04510 D.F., M\'exico.}

\date{\today}


\begin{abstract}
We computed flat rotation curves from scalar-tensor theories in their weak field limit. 
Our model, by construction, fits a flat rotation profile for 
velocities of stars. As a result, the form of the scalar field potential and DM distribution
in a galaxy are determined. By taking into account the constraints
for the fundamental parameters of the theory $(\lambda, \, \alpha)$,
it is possible to obtain analytical results for the density profiles.  For positive and negative  values of $\alpha$, 
the DM matter profile is as cuspy as NFW's.

\end{abstract}


\maketitle

\section{Introduction}
Recently, we have worked on  some of the effects that general scalar-tensor theories
(STT) of gravity yield 
on astrophysical scales \cite{Ro01,RoCe04,RoCePeTlCa05}.  
By taking the  weak field 
limit of  STT, three parameters  appear \cite{PiOb86}:  the Newtonian constant at infinity
 $G_{\infty}$, 
a Compton length-scale ($\lambda=\hbar/m c $) coming from the effective mass
of the Lagrangian 
potential, and the  strength of the new scalar gravitational interaction ($\alpha$).   
For point-like 
masses, the new Newtonian potential is of a Yukawa type \cite{PiOb86,FiTa99} 
and, in general, analytical expressions can be found for spherical \cite{RoCe04} 
and axisymmetric systems \cite{RoCePeTlCa05}.   Using these solutions we build
 a galactic model that
is consistent with a flat rotation velocity profile which resembles the 
observations \cite{SoRu01}.  In the 
 present short contribution, we outline a way to construct such a galactic model using  STT.     

\section{STT and constraints} \label{st-constraints}
A typical spiral galaxy consists of a disk, a bulge,  and a dark matter (DM) halo.   A real halo is roughly spherical 
in shape and contains most of the matter, up to 90\%, of the system.  Thus, being the halo the main component, 
our galactic model will then consist of a spherical system of unknown DM. Therefore, in order compare with observations we assume that test 
particles --stars and dust-- follow DM particles. To construct our halo model, we take the weak field limit of general STT 
and consider a spherically symmetric fluctuation of the scalar field around some fixed background field.  The 
STT is then characterized by a background value of the scalar field, $\langle \phi \rangle$, a Compton 
lengthscale, $\lambda$, and a strength $\alpha$ \cite{PiOb86}.  For point-like sources, the new Newtonian 
potential is well known to be  of a Yukawa type \cite{FiTa99}:
\begin{equation} \label{phi_New} 
\Phi_N = -\frac{1}{\langle \phi \rangle} \frac{M}{r} (1+ \alpha \,
\rm{e}^{-r/\lambda}) \, .
\end{equation}
If one fixes the background field to be the inverse of the Newtonian constant, $\langle \phi \rangle= G_{N}^{-1}$ , then 
for $r \gg \lambda$ the new Newtonian  potential coincides with the standard Newtonian one. Given 
this, for $r \ll \lambda$ one finds deviations of the order of $(1+\alpha)$ 
to the Newtonian dynamics. This setting is, however, very constrained by local (solar system) deviations of the 
Newtonian force, in order for  $\alpha$ to be less than $10^{-10}$ \cite{FiTa99}.  Alternatively, one can choose 
the setting $\langle \phi \rangle= G_{N}^{-1}(1+\alpha)$ and the new potential coincides with the 
Newtonian for $r \ll \lambda$ and deviates by $1/(1+\alpha)$ for $r \gg \lambda$. If one thinks of a galactic 
system, physical scales are around the tens of kiloparsecs --and so is our typical $\lambda$--, then for distances 
bigger than this, one expects constrictions, e.g. from cluster dynamics or cosmology.   Several authors  
\cite{constraints} have considered these deviations, giving a rough estimate within the 
range  of $-1 < \alpha \lesssim 5$.  For example, the value of $\alpha=-0.5$ yields an asymptotic growing factor of $2$ in
$G_N$, whereas the value of $\alpha=3.0$ reduces $G_N$ asymptotically by one fourth.
  
\section{The galactic model} \label{gal-model}
\begin{figure}
\center{\includegraphics[width=88mm]{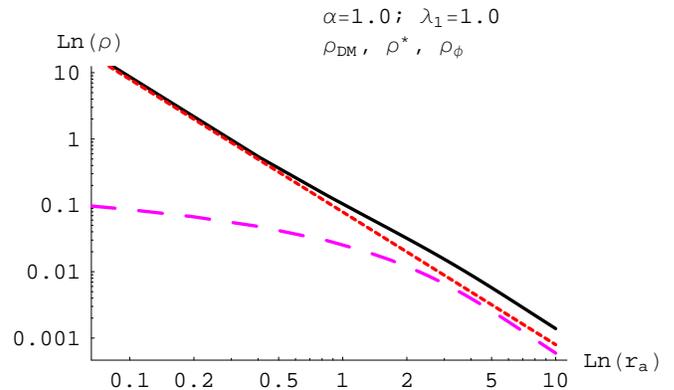}}
\caption{All density profiles as a function of the radius are shown in a log-log scale.  The plots were made with
$\lambda=1.0$ and $\alpha=1.0$.  $\rho_{DM}$ (solid line), $\rho^{*} $ (short-dashed line) and $ \rho_{\bar \phi} $ (long-dashed line).} 
\label{all-densities}
\end{figure}

\begin{figure}
\center{\includegraphics[width=88mm]{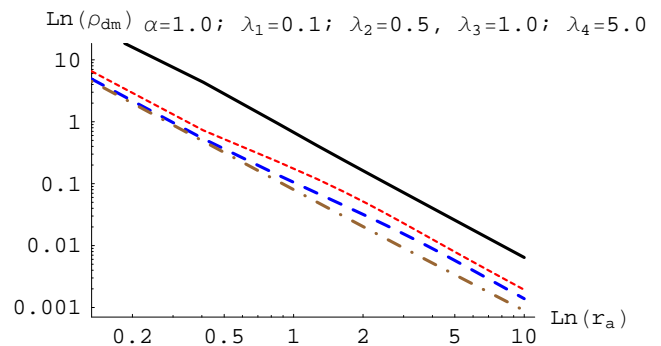}}
\caption{DM density profiles as a function of the radius are shown in log-log scale.  The plots were made with
$\alpha=1.0$ and $\lambda_{1}=0.1$ (solid line), $\lambda_{2}=0.5$ (short-dashed line), $\lambda_{3}=1.0$ (long-dashed line), and 
$\lambda_{4}=5.0$ (dashed-dotted line). }
\label{various-lambda}
\end{figure}

For a general density distribution the equations governing the weak energy 
(Newtonian limit) of STT are \cite{He91,RoCe04}: 
\begin{eqnarray}
 \nabla^2 \Phi_{N} &=& \frac{G{_N}}{1+\alpha} \, \left[
4\pi \rho - \frac{1}{2} \nabla^2 \bar{\phi} \right] \; ,
\label{pares_eq_h00}\\
  \nabla^2 \bar{\phi} - m^2 \bar{\phi} &=&
-  8\pi \alpha\rho \; , \label{pares_eq_phibar}
\end{eqnarray}
where $\bar{\phi} $ is the scalar field  fluctuation and $\rho$ is the density 
distribution which contains baryons, DM particles, or other types of matter. For definiteness 
we can think of DM, but its profile is unknown yet. This will  be 
determined by setting the dynamics of test particles, i.e., by demanding 
the new Newtonian potential to be the one that solves the rotation curves of test 
particles --stars and dust. That is, we require  
\begin{equation} \label{ph-rot-prof} 
v_{c}^{2} = r \frac{d\Phi_N}{dr} =  {\rm const.} \, ,
\end{equation}
where this constant is chosen to fit rotation velocities in spirals \cite{SoRu01}; this is the simplest model. We then 
proceed to solve for $\Phi_N$, and its solution is: 
\begin{equation} \label{ph-N}  
\Phi_N =  v_{c}^{2} \, \ln(r)
\, .
\end{equation}
Substituting this result into the original system,
(\ref{pares_eq_h00}, \ref{pares_eq_phibar}), gives
\begin{equation} \label{rho_sol}  
\rho \equiv \rho_{DM} =    \frac{v_{c}^{2}}{4 \pi G_N r^2} + \frac{m^2}{8 \pi (1+\alpha)} \bar{\phi} \, . 
\end{equation}
For our convenience, we define the following densities: $ \rho^{*} \equiv \frac{v_{c}^{2}}{4 \pi G_N r^2}  $ and 
$ \rho_{\bar \phi}  \equiv \frac{m^2}{8 \pi (1+\alpha)}  \bar{\phi} $,  where  
$\rho^{*}$ is the density that the system would have to achieve flat rotation curves if system would be treated solely with Newtonian physics,  
and $ \rho_{\bar \phi}$ is the contribution of the scalar field, which should be obtained by integrating  the equation:
\begin{equation} \label{nablaphi}
\nabla^2 \bar{\phi} - \frac{m^2}{1+\alpha} \bar{\phi} = -  8 \pi \alpha \rho^{*}  \, .
\end{equation}
By comparing (\ref{pares_eq_phibar}) with (\ref{nablaphi}), it seems natural to identify  
$m = m^{*}  \sqrt{1+\alpha}$  to convert (\ref{nablaphi}) to a type 
(\ref{pares_eq_phibar}) equation, now for $m^{*}$ and $\rho^{*}$.  The solution is therefore given 
by \cite{RoCe04}:
\begin{eqnarray}\label{pares_eq_finalphi} 
\bar{\phi}(r) &=& 8 \pi \alpha a^{2} \left[  \frac{\rm{e}^{-r_{a}/ \lambda^{*}_{a}}}{r_a}  
\int_0^{r_{a}} dx \; x \;  \sinh(x/ \lambda^{*}_{a})   \rho^{*}(x)   \right. \nonumber \\
&& \left. 
+ \frac{\sinh(r/ \lambda^{*}_{a})}{r_a}  \int_{r_{a}}^{R_{a}} dx \; x \; \rm{e}^{-x/ \lambda^{*}_{a}}  
\rho^{*}(x) \right]  \, ,
\end{eqnarray}
where $\lambda^{*}  \equiv \lambda \sqrt{1+\alpha}$ and  $a$   is a length scale of  the 
spherical system, e.g. related to the distance at which stars possess  flat rotation curves; 
the size of the halo is denoted by $R$.  Once (\ref{pares_eq_finalphi}) is 
solved, its solution is substituted into  (\ref{rho_sol}) to 
find the DM distribution, $\rho_{DM}$.  The procedure  outlined here is straightforward.  Thus, 
using our given $\rho^{*}$ the above  expression can analytically be 
integrated to have 
\begin{eqnarray} \label{phi-bar-sol} 
\bar{\phi}(r) &=&  \frac{2  \alpha \, a^{2} v_{c}^{2} }{G_{N} r_{a}} 
\left[  \rm{e}^{-\frac{r_a}{\lambda^{*}_a }} \,   
\rm{sinhIntegral}\left(\frac{r_a}{\lambda^{*}_a }\right)
\nonumber \right. \\  && \left.
-\rm{sinh}\left(\frac{r_a}{\lambda^{*}_a}\right) \,  
\rm{expIntegralEi}\left(-\frac{r_a}{\lambda^{*}_a } \right)
\right]  \, . 
\end{eqnarray}
Thus, the DM profile becomes
\begin{eqnarray} \label{rho-dm-sol} 
\rho_{DM}  &=& \frac{v_{c}^{2}}{4 \pi G_{N}} 
\left\{  \frac{1}{a^{2} r_a^2} 
\right. \nonumber \\ &&
+  \frac{\alpha}{1+\alpha}  \frac{1}{\lambda_{a}^{2}\, r_{a} } 
 \left[   \rm{e}^{-\frac{r_a}{\lambda^{*}_a }} \,   
 \rm{sinhIntegral}\left(\frac{r_a}{\lambda^{*}_a }\right)
 \right. \nonumber  \\ 
&&\left. \left. 
- \rm{sinh}\left(\frac{r_a}{\lambda^{*}_a}\right) \,  
\rm{expIntegralEi}\left(-\frac{r_a}{\lambda^{*}_a } \right)
\right]  \right\} \, . 
\end{eqnarray}
Following, we plot in figure \ref{all-densities} the density profiles
  $\rho_{DM}$, $\rho^{*} $ and $ \rho_{\bar \phi} $ for $\lambda=1$, $\alpha=1$.
The main contribution to $\rho_{DM}$ comes from $\rho^{*} $, which is an inverse squared function of the radius. The DM profile is therefore cuspy
near the galactic centre, similar to the NFW's obtained from simulations \cite{Na96-97}.   We have computed the best fits to these curves for $r\ll a$, 
given by   $\rho_{DM} \sim \frac{0.08}{r^{1.99}}$ and $ \rho_{\bar \phi}  \sim \frac{0.06}{r^{0.21}}$.  On the other hand, the best fits for $r\gg a$ are 
$\rho_{DM} \sim \frac{0.16}{r^{2.05}}$ and $ \rho_{\bar \phi}  \sim \frac{0.08}{r^{2.12}}$.  In figure \ref{various-lambda} we plot the DM profile 
for various $\lambda$ values, resulting in small changes in the slope.  For negative $\alpha$ values, (\ref{phi-bar-sol}) becomes  negative, 
but the DM profile does not become shallower.  We have 
computed the case $\lambda=1.0$, $\alpha=-0.5$  for $r\ll a$ and found $\rho_{DM} \sim \frac{0.08}{r^{2.00}}$, while for $r\gg a$ 
we obtained $\rho_{DM} \sim \frac{0.02}{r^{1.83}}$. 


\section{Conclusions}

We have  constructed a spherically galactic halo model in which we fit the new gravitational 
potential to match a flat rotation profile for test particles --stars and dust.  Once this is done, one computes the scalar
field fluctuation and DM profiles resulting from this potential.  We have found analytical results
for our model, Eqs.(\ref{phi-bar-sol}) and (\ref{rho-dm-sol}).   
These equations contain the information of the
scalar field ($\lambda$, $\alpha$), which determine the DM
profile. We have plotted our results for $\alpha=1.0$ and various $\lambda$s. These profiles have  an 
inner region as cuspy as the NFW's. Negative values of $\alpha$ do not yield a shallower DM profile neither.  This 
result may however not surprise since we are not resolving the inner structure of the rotation velocity curves. We
only considered its flat asymptotic behaviour; in a forthcoming paper we will study it. 

\bigskip
{\it Acknowledgements: }
This work was supported by CONACYT grant numbers 44917 and U47209-F.


\end{document}